\documentclass{PoS}
\usepackage{amsmath}
\usepackage{amssymb}
\usepackage{graphicx}
\usepackage{fontenc}
\usepackage{times}
\usepackage{mathptmx}

\title{Exploring Gauge-Invariant Vacuum Wave Functionals for Yang-Mills
Theory}
\author{\speaker{Hilmar Forkel} \\
Institut f\"ur Physik, Humboldt-Universit\"at zu Berlin, D-12489 Berlin,
Germany \\
E-mail: \email{forkel@physik.hu-berlin.de}}

\ShortTitle{Exploring Gauge-Invariant Vacuum Wave Functionals 
for Yang-Mills Theory}
\abstract{
We study gauge-invariant approximations to the Yang-Mills vacuum 
wave functional in which asymptotic freedom and a detailed
description of the infrared dynamics are encoded through 
squeezed core states. After variationally optimizing these trial 
functionals, dimensional transmutation, gluon condensation and a 
dynamical mass gap of the expected magnitude emerge transparently. 
The dispersion properties of the soft gauge modes are modified
by higher-gradient interactions and suggest a negative differential 
color resistance of the Yang-Mills vacuum. Casting the soft-mode 
dynamics into the 
form of an effective action for gauge-invariant collective fields, 
furthermore, allows to identify novel infrared degrees of freedom. 
The latter are gauge-invariant saddle-point fields which summarize 
dominant and 
universal contributions from various gauge-field orbits to all 
amplitudes. Their analysis provides new insights into how the 
vacuum gluon fields generate gauge-invariant excitations. 
Examples include a dynamical size stabilization mechanism for 
instantons and merons, a gauge-invariant representation of their 
effects as well as a new physical interpretation for Faddeev-Niemi 
knots.}
\FullConference{International Workshop on QCD Green's Functions, Confinement and
Phenomenology\\ 5-9 September 2011\\ Trento, Italy}

\begin{document}

\section{Introduction}

The Hamiltonian formulation of Yang-Mills (YM) theory in the Schr\"{o}dinger
picture, although not particularly efficient in the perturbative domain,
offers considerable benefits when addressing nonperturbative issues. Among
its attractive features are the explicit representation of the vacuum state
which invokes quantum mechanical intuition \cite{fey81}, the ability to
treat genuine real-time problems (including non-equilibrium processes) as
well as the transparent description of topological effects \cite{jac90}. In
particular, however, it makes gauge theories accessible to a variational
treatment \cite{fey81,varYM}, i.e. to one of the few approximation schemes
currently available for strongly coupled quantum field theories.

Variational calculations in Yang-Mills theories are often performed in a
fixed gauge, most notably in Coulomb gauge \cite{sze04}. In the following we
will report on our complementary explorations \cite{for10} of a manifestly
gauge-invariant formulation of the variational problem \cite{kog95}. This
framework renders fundamental infrared (IR) physics, including dimensional
transmutation and the generation of a mass gap, particularly transparent.
Moreover, it preserves the full topological structure of the gauge group.
The latter is particularly relevant since topological properties are likely
robust enough to survive limitations of the restricted trial functional
basis which keeps the approach analytically manageable. Another attractive
feature of the gauge-invariant formulation is that the infrared dynamics can
be re-expressed in terms of gauge-invariant collective fields which subsume
contributions from whole gauge-field orbit families. After performing an IR
improved variational analysis \cite{for10}, we will make use of this feature
to identify gauge-invariant and universal IR degrees of freedom of the gauge
dynamics \cite{for06}. More details can be found in Refs. \cite{for10,for06}.

\section{Gauge-invariant vacuum wave functionals}

\label{ginv}

Starting from an approximate and hence typically gauge-dependent\
\textquotedblleft core\textquotedblright\ functional $\psi _{0}\left[ \vec{A}%
\right] $ of the static gauge fields (i.e. of half of the canonical
variables), we impose gauge invariance by averaging over the gauge group.
The result is a trial vacuum wave functional (VWF) of the form%
\begin{equation}
\Psi _{0}\left[ \vec{A}\right] =\sum_{Q\in Z}e^{iQ\theta }\int D\mu \left[
U^{\left( Q\right) }\right] \psi _{0}\left[ \vec{A}^{U^{\left( Q\right) }}%
\right] =:\int DU\psi _{0}\left[ \vec{A}^{U}\right]   \label{ginvvwf}
\end{equation}%
where $d\mu $ is the Haar measure of the SU$\left( N_{c}\right) $ gauge
group, $Q$ the topological (homotopy) charge of the group element $U^{\left(
Q\right) }$, and $\theta $ the vacuum angle. Since the vacuum wave
functional is nodeless \cite{fey81}, one may write $\psi _{0}\left[ \vec{A}%
\right] =\mathcal{N}^{-1}\exp \left( -\Phi \left[ \vec{A}\right] \right) $
and expand the real functional $\Phi $ into a power series. The constant
term is absorbed into $\mathcal{N}$ and the term linear in $A$ is generally
discarded (coherent gluon vacuum states are known to be unstable \cite{leu81}%
).

The next term is quadratic in $A$ and plays several crucial roles. First, it
removes the ambiguity in $\Psi _{0}$ \cite{zar98} due to the invariance of
the Haar measure in Eq. (\ref{ginvvwf}) under group transformations.
Furthermore, this term can incorporate asymptotic freedom and thus render
the VWF exact in the ultraviolet. Finally, and from the practical
perspective most importantly, functionals resulting from a quadratic term
can be integrated over $A$ analytically. Hence one generally truncates the
series for $\Phi $ after the quadratic term, which leads to the
\textquotedblleft squeezed\textquotedblright\ core\ functional 
\begin{equation}
\psi _{0}^{\left( G\right) }\left[ \vec{A}\right] =\frac{1}{\mathcal{N}_{G}}%
\exp \left[ -\frac{1}{2}\text{ }\int d^{3}x\int d^{3}yA_{i}^{a}\left( \vec{x}%
\right) G_{ij}^{-1ab}\left( \vec{x}-\vec{y}\right) A_{j}^{b}\left( \vec{y}%
\right) \right]  \label{ga}
\end{equation}%
with the normalization factor $\mathcal{N}_{G}^{-1}=\left[ \det \left(
G/2\right) \right] ^{-1/4}$ and a real \textquotedblleft
covariance\textquotedblright\ $G^{-1}$.

\subsection{Gluon dispersion: asymptotic freedom and IR generality}

\label{guv}

We now have to specify the properties of the function $G^{-1}$ in the trial
functional family (\ref{ga}). Translational invariance was already
anticipated in Eq. (\ref{ga}). We will further restrict ourselves to a
purely transverse covariance with the Fourier transform \cite{kog95} 
\begin{equation}
G_{ij}^{-1,ab}\left( k\right) =\delta _{ij}\delta ^{ab}G^{-1}\left( k\right)
\label{gspat}
\end{equation}%
(cf. Ref. \cite{for10} for a discussion of this choice 
and Refs. \cite{dia98,bro98} for the
impact of longitudinal contributions). The normalizability of physical wave
functionals then demands $G^{-1}\left( k\right) >0$ and further ensures
vacuum stability and a positive energy spectrum. In order to implement the
correct UV behavior, we factorize the core functionals (\ref{ga}) as $\psi
_{0}^{\left( G\right) }\left[ \vec{A}\right] =\psi _{0}^{\left( G_{<}\right)
}\left[ \vec{A}_{<}\right] \psi _{0}^{\left( G_{>}\right) }\left[ \vec{A}_{>}%
\right] $ by splitting the $\vec{k}$ integration domain in their
exponentials into soft/hard momentum regions with $\left\vert \vec{k}%
\right\vert \gtrless \mu $. The separation scale $\mu $ will be determined
below. Asymptotic freedom requires $G$ to approach the non-interacting,
massless static vector propagator $G_{0}\left( k\right) =1/k$ for $%
k\rightarrow \infty $. As long as $\mu \gg \Lambda _{\text{YM}}$ (where $%
\Lambda _{\text{YM}}$ is the Yang-Mills scale) perturbative hard-mode
corrections remain small, which allows us to approximate 
\begin{equation}
G_{>}^{-1}\left( k\right) =G_{0}^{-1}\left( k\right) =k.  \label{gm1uv}
\end{equation}%
The unknown IR covariance $G_{<}^{-1}\left( k\right) $, on the other hand,
will be determined variationally. In Ref. \cite{kog95} the minimal
one-parameter trial function $G_{<,\text{KK}}^{-1}\left( k\right) =\mu $ was
adopted. We have implemented a far more comprehensive parametrization \cite%
{for06}, based on the under reasonable analyticity assumptions general and
controlled gradient expansion%
\begin{equation}
G_{<}^{-1}\left( \vec{x}-\vec{y}\right) =m_{g}\left[ 1+c_{1}\frac{%
\partial _{x}^{2}}{\mu ^{2}}+c_{2}\left( \frac{\partial _{x}^{2}}{\mu ^{2}}%
\right) ^{2}+c_{3}\left( \frac{\partial _{x}^{2}}{\mu ^{2}}\right) ^{3}+...%
\right] \delta _{<}^{3}\left( \vec{x}-\vec{y}\right) .  \label{gm1}
\end{equation}%
Eq. (\ref{gm1}) can be efficiently truncated to maintain an analytically
manageable trial basis for the soft-mode physics. Besides $\mu $, the
variational parameter space now contains the IR gluon mass\ $m_{g}>0$
and a few low-momentum constants $c_{i}$ which characterize dispersive gluon
properties in the vacuum. The regularized delta function $\delta
_{<}^{3}\left( \vec{x}-\vec{y}\right) :=\int d^{3}k/\left( 2\pi \right)
^{3}\theta \left( \mu ^{2}-\vec{k}^{2}\right) e^{i\vec{k}\left( \vec{x}-\vec{%
y}\right) }$ encodes the slow variation $\left\Vert \partial
A_{<}\right\Vert /\left\Vert A_{<}\right\Vert \leq \mu $ of the soft modes
and ensures that the higher-order terms in Eq. (\ref{gm1}) are
parametrically suppressed.

As a consequence of $G^{-1}\left( k\right) >0$, the low-momentum constants
are subject to the bounds $c_{1}<1,$ $c_{2}>-1$, etc. (for $m_{g}>0$%
). Requiring continuity of $G^{-1}\left( k\right) $ at the matching point $%
k=\mu $, furthermore, fixes $m_{g}$ as a function of the other
variational parameters. When truncating\ to $c_{i\geq 2}=0$, for example,
one has%
\begin{equation}
m_{g}\left( \mu ,c_{1}\right) =\frac{\mu }{1-c_{1}}.  \label{mgc}
\end{equation}%
Note that the requirement of a non-negative IR gluon mass restricts the $%
c_{1}$ domain to $c_{1}<1$, in agreement with the above bound from $%
G^{-1}\left( k\right) >0$. The VWF (\ref{ginvvwf})\ together with the core
functional (\ref{ga}) and the covariance (\ref{gm1uv}), (\ref{gm1})
(possibly with perturbative corrections)\ appears to be the
\textquotedblleft richest\textquotedblright\ gauge-invariant\ trial
functional\ family whose matrix elements can be calculated analytically by
currently available techniques.

\section{Variational analysis}

One the basis of the trial functional family (\ref{ginvvwf}) discussed
above, the variational analysis amounts to minimizing the expectation value 
\begin{equation}
\left\langle \mathcal{H}\left( A,E\right) \right\rangle =\frac{\int D\vec{A}%
\Psi _{0}^{\ast }\left[ \vec{A}\right] \mathcal{H}\left( \vec{A}^{a},\vec{E}%
^{a}\right) \Psi _{0}\left[ \vec{A}\right] }{\int D\vec{A}\Psi _{0}^{\ast }%
\left[ \vec{A}\right] \Psi _{0}\left[ \vec{A}\right] }  \label{hexp}
\end{equation}
of the Yang-Mills Hamiltonian density 
\begin{equation}
\mathcal{H}=\frac{1}{2}\left( E_{i}^{a}E_{i}^{a}+B_{i}^{a}B_{i}^{a}\right)
\label{HYM}
\end{equation}
(in temporal gauge, with $\vec{E}^{a}=i\delta /\delta \vec{A}^{a}$) with
respect to the parameters appearing in $G^{-1}$. After inserting the wave
functional (\ref{ginvvwf}) into Eq. (\ref{hexp}) and interchanging the order
of integration over fields and group elements, the gauge invariance of the $%
\vec{A}$ integral allows to factor out a gauge group volume. Eq. (\ref{hexp}%
) can thus be rewritten as 
\begin{equation}
\left\langle \mathcal{H}\left( A,E\right) \right\rangle =\frac{\int DU\int D%
\vec{A}\psi _{0}\left[ \vec{A}^{U}\right] \mathcal{H}\left( \vec{A}^{a},%
\frac{i\delta }{\delta \vec{A}^{a}}\right) \psi _{0}\left[ \vec{A}\right] }{%
\int DU\int D\vec{A}\psi _{0}\left[ \vec{A}^{U}\right] \psi _{0}\left[ \vec{A%
}\right] }
\end{equation}%
(where $DU$ is the functional SU$\left( N_{c}\right) $ measure as defined in
Eq. (\ref{ginvvwf})). After evaluating the functional derivatives contained
in $\mathcal{H}$, the Gaussian integration over $A$ can be performed
exactly, resulting in%
\begin{equation}
\left\langle \mathcal{H}\right\rangle =\frac{\int DU\left\langle
\left\langle \left\langle \mathcal{H}\right\rangle \right\rangle
\right\rangle \exp \left\{ -\Gamma _{b}\left[ U\right] \right\} }{%
\int DU\exp \left\{ -\Gamma _{b}\left[ U\right] \right\} }
\end{equation}%
where we introduced the notation 
\begin{equation}
\left\langle \left\langle \left\langle \vec{A}...\vec{A}...\vec{E}...\vec{E}%
\right\rangle \right\rangle \right\rangle \exp \left\{ -\Gamma _{b}%
\left[ U\right] \right\} \equiv \int D\vec{A}\psi _{0}\left[ \vec{A}^{U}%
\right] \vec{A}...\vec{A}...\frac{i\delta }{\delta \vec{A}}...\frac{i\delta 
}{\delta \vec{A}}\psi _{0}\left[ \vec{A}\right]
\end{equation}%
for matrix elements between $U$-rotated and unrotated core VWFs. The above
expression defines, in particular, the effective bare action $\Gamma _{b}
\left[ U\right] =-\ln \int D\vec{A}\,\ \psi _{0}^{\ast }\left[ \vec{A}^{U}%
\right] \psi _{0}\left[ \vec{A}\right] $ which describes dynamical
correlations generated by the gauge projection. This action gathers all
those gauge-field contributions to the generating functional whose
approximate vacua $\psi _{0}$ at $t=\pm \infty $ differ by a relative gauge
orientation $U$. Hence the gauge-invariant \textquotedblleft
variable\textquotedblright\ $U$ represents the contributions from all such
gluon field orbits to the vacuum overlap. Explicitly, one finds \cite{for10} 
\begin{equation}
\Gamma _{b}\left[ U\right] =\frac{1}{2g_{\text{b}}^{2}}\int
d^{3}x\int d^{3}yL_{i}^{a}\left( \vec{x}\right) \mathcal{D}^{ab}\left( \vec{x%
}-\vec{y}\right) L_{i}^{b}\left( \vec{y}\right)  \label{gamb}
\end{equation}%
with $L_{i}=U^{\dagger }\partial _{i}U=:L_{i}^{a}\lambda ^{a}/\left(
2i\right) $ and $\mathcal{D}^{ab}=\left[ \left( G+G^{U}\right) ^{-1}\right]
^{ab}\simeq \frac{1}{2}G^{-1}\delta ^{ab}+...$ where $G^{U}=G^{ab}\left( 
\vec{x}-\vec{y}\right) $ $U^{\dagger }\left( \vec{x}\right) \left( \lambda
^{a}/2\right) U\left( \vec{x}\right) \otimes U\left( \vec{y}\right) \left(
\lambda ^{b}/2\right) U^{\dagger }\left( \vec{y}\right) $.

After splitting $U\left( \vec{x}\right) =U_{<}\left( \vec{x}\right)
U_{>}\left( \vec{x}\right) $ with $U_{>}\left( \vec{x}\right) =\exp \left(
-ig\phi ^{a}\left( \vec{x}\right) \lambda ^{a}/2\right) $ into hard- and
soft-mode contributions and integrating over the hard modes $\phi ^{a}$
perturbatively \cite{kog95}, furthermore, one arrives at%
\begin{equation}
\left\langle \mathcal{H}\right\rangle =\frac{\int DU_{<}\int D\phi
\left\langle \left\langle \left\langle \mathcal{H}\right\rangle
\right\rangle \right\rangle \exp \left\{ -\Gamma _{b}\left[ \phi
,U_{<}\right] \right\} }{\int DU_{<}\int D\phi \exp \left\{ -\Gamma _{b}
\left[ \phi ,U_{<}\right] \right\} }.
\end{equation}%
With the additional definition%
\begin{equation}
\left\langle \left\langle \mathcal{O}\right\rangle \right\rangle \exp
\left\{ -\Gamma _{<}\left[ U_{<}\right] \right\} :=\int D\phi \left\langle
\left\langle \left\langle \mathcal{O}\right\rangle \right\rangle
\right\rangle \exp \left\{ -\Gamma _{b}\left[ \phi ,U_{<}\right]
\right\} ,  \label{o2}
\end{equation}%
which contains the effective soft-mode action 
\begin{equation}
\Gamma _{<}\left[ U_{<}\right] :=-\ln \int D\phi \exp \left\{-\Gamma _{b}
\left[ \phi ,U_{<}\right] \right\}  \label{gsm0}
\end{equation}
(i.e. the RG evolved bare action (\ref{gamb})), we can finally rewrite the
matrix element (\ref{hexp}) solely in terms of the $U_{<}$ field dynamics,
i.e.%
\begin{equation}
\left\langle \mathcal{H}\right\rangle =\frac{\int DU_{<}\left\langle
\left\langle \mathcal{H}\right\rangle \right\rangle \exp \left\{ -\Gamma _{<}%
\left[ U_{<}\right] \right\} }{\int DU_{<}\exp \left\{ -\Gamma _{<}\left[
U_{<}\right] \right\} }.  \label{hsmvev}
\end{equation}
Since the \textquotedblleft reduced\textquotedblright\ (i.e., fixed $U_{<}$)
matrix element $\left\langle \left\langle \mathcal{H}\right\rangle
\right\rangle $ is a nonlocal functional of the soft modes $U_{<}$, the
evaluation of Eq. (\ref{hsmvev}) amounts to calculating (equal-time)
soft-mode correlation functions \cite{for10}.

\subsection{Vacuum phases}

\begin{figure}[tbp]
\begin{center}
\includegraphics[width=.5 \textwidth]{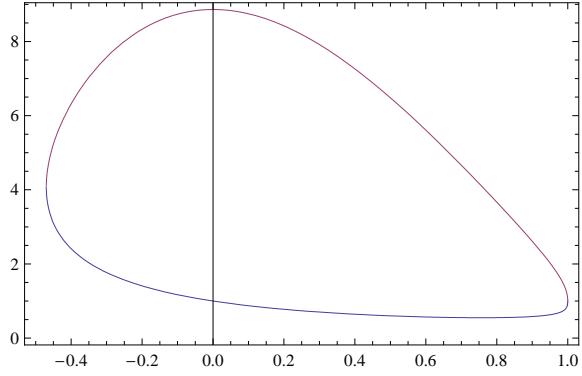}
\end{center}
\caption{The vacuum phase diagram. Inside the plotted phase boundary $%
\protect\mu _{c}\left( c_{1}\right) /\Lambda_{\text{YM}}$ the theory is in
its strongly-coupled disordered phase. (The underlying approximations are
reliable for $\protect\mu\gtrsim4\Lambda_{\text{YM}}$ and $c_{1}\in\left\{
-0.5,0.5\right\} $).}
\label{cl}
\end{figure}

In integrals over the $U_{<}$ fields (such as those in Eq. (\ref{hsmvev}))
the unitarity constraint $U_{<}^{\dagger }U_{<}=1$ can be resolved by
inserting a delta functional which is then written as an additional integral
over a hermitean auxilary field $\Sigma $. In Eq. (\ref{hsmvev}) the
integration over the then unconstrained $U_{<}$ becomes Gaussian and can be done
analytically. In the mean-field approximation, the expression for $%
\left\langle \left\langle \mathcal{H}\right\rangle \right\rangle $ is then
evaluated at the saddle point $\bar{\Sigma}=:\left( \mu \bar{\xi}\right)
^{2} $ of the $\Sigma $ integral, i.e. at the minimal-action solution of the
gap equation 
\begin{equation}
\left\langle U_{<,AB}^{\dagger }\left( \vec{x}\right) U_{<,BC}\left( \vec{x}%
\right) \right\rangle =\delta _{AC}  \label{geq2}
\end{equation}%
which reintroduces unitarity in the mean. After adopting the one-loop
Yang-Mills coupling $\gamma \left( \mu \right) =g_{\text{YM}}^{2}\left( \mu
\right) N_{c}/\pi ^{2}\overset{N_{c}=3}{=}24/\left( 11\ln \mu /\Lambda _{%
\text{YM}}\right) $, the solutions $\bar{\xi}$ of Eq. (\ref{geq2}) depend on
two variational parameters, the RG scale $\mu \geq 0$ and $c_{1}<1$. The
critical line $\mu _{c}\left( c_{1}\right) $, i.e. the parameter
subspace where the (dis-)order parameter $\bar{\xi}\left( \mu _{c}
\left( c_{1}\right) ,c_{1}\right) $ vanishes and the phase transition takes
place, can be found analytically\ as the combination of the two curves 
\begin{equation}
\frac{\mu _{c,1,2}\left( c_{1}\right) }{\Lambda _{\text{YM}}}=\exp %
\left[ \frac{48}{11}\frac{\left( 1-c_{1}\right) \left[ 1-\tilde{\imath}%
\left( c_{1}\right) \right] \left[ 1+\left( 1-c_{1}\right) \tilde{\imath}%
\left( c_{1}\right) \right] }{\left( 1-c_{1}\right) \tilde{\imath}\left(
c_{1}\right) \pm \sqrt{5\tilde{\imath}^{2}\left( c_{1}\right) \left(
1-c_{1}\right) ^{2}-4\left( 1-c_{1}\right) \left[ 1-c_{1}\tilde{\imath}%
\left( c_{1}\right) \right] }}\right]  \label{mu}
\end{equation}%
($\tilde{\imath}\left( c_{1}\right) :=\mathrm{arctanh}\sqrt{c_{1}}/\sqrt{%
c_{1}}$). We plot this closed phase boundary in Fig. \ref{cl}. It limits the
parameter ranges to $0.5\lesssim \frac{\mu _{\text{c}}}{\Lambda _{\text{YM}}}%
\lesssim 8.86$ and $-0.48\lesssim c_{1}<1\ $and thus prevents the
minimal-energy solution $\bar{\xi}^{\ast }$ from attaining unacceptably
large values of $\mu $ and $\left\vert c_{1}\right\vert $. Nonzero solutions
of the gap equation exist only when the gauge coupling exceeds a critical
value, i.e. for $g^{2}\left( \mu \right) >g_{\text{c}}^{2}\left(
c_{1}\right) ,$ as expected on physical grounds. The (dis-)order parameter
goes to zero continuously, furthermore, i.e. the disorder-order transition
is of second order (which may be an artefact of the mean-field approximation 
\cite{for10}).

\subsection{Vacuum energy density}

\label{vend}

\begin{figure}[tbp]
\begin{center}
\includegraphics[width=.6\textwidth]{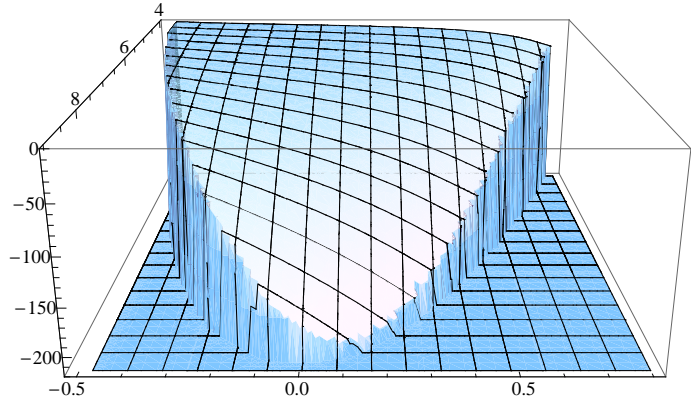}
\end{center}
\caption{The energy density $\bar{\protect\varepsilon}\left(\protect\mu%
,c_{1}\right)$ of the vacuum field solution $\bar{\protect\xi}\left(\protect%
\mu,c_{1}\right) $ in the disordered phase. (The plot shows the parameter ranges $\protect\mu\in\left\{4,9\right\} \Lambda_{\text{YM}}$ and $c_{1}\in\left\{-0.5,0.8\right\}.$) Note
the minimum of the energy surface at $c_{1}\simeq0.15$.}
\label{ed}
\end{figure}

Working with the Poincar\'{e}-invariant trial states (\ref{ginvvwf}) and
taking only one-loop corrections from the hard modes into acccount, it is
sufficient to regularize Eq. (\ref{hexp}) by a momentum cutoff $\Lambda _{%
\text{UV}}$ \cite{kog95}. Separating the complete vacuum energy density $%
\varepsilon =E/V=\left\langle \mathcal{H}_{\text{YM}}\right\rangle $ into
hard and soft contributions,%
\begin{equation}
\varepsilon \left( \mu ,c_{1},\zeta ;\bar{\xi}\right) =\left\langle \mathcal{%
H}_{\text{YM}}\right\rangle =\varepsilon _{>}\left( \mu \right) +\varepsilon
_{<}\left( \mu ,c_{1},\zeta ;\bar{\xi}\right) ,  \label{etot}
\end{equation}%
($\zeta \equiv m_{g}/\mu $) the cutoff dependence resides solely in%
\begin{equation}
\varepsilon _{>}\left( \mu \right) =\frac{N_{c}^{2}-1}{8\pi ^{2}}\left(
\Lambda _{\text{UV}}^{4}-\mu ^{4}\right) .  \label{epsh}
\end{equation}%
As expected, this is the (regularized) zero-point energy density of two
transverse, \emph{massless} vector modes in the adjoint representation
of SU($N_{c}$) with energy $\omega \left( k\right)
=k$. Simple normal-ordering thus subtracts the $\Lambda _{\text{UV}}$
dependent term. For $c_{i\geq 2}=0$ one then finds the total energy density $%
\bar{\varepsilon}\left( \mu ,c_{1}\right) :=$ $\varepsilon \left( \mu
,c_{1},\zeta _{\text{ct}}\left( c_{1}\right) ;\bar{\xi}\left( \mu
,c_{1}\right) \right) $ in the disordered phase as%
\begin{align}
\bar{\varepsilon}\left( \mu ,c_{1}\right) =& -\frac{N_{c}^{2}}{4\pi ^{2}}\mu
^{4}\left[ \frac{4c_{1}^{3}+10c_{1}^{2}-50c_{1}+30}{30c_{1}^{2}\left(
1-c_{1}\right) }-\frac{1-c_{1}}{c_{1}^{2}}\frac{\mathrm{arctanh}\sqrt{c_{1}}%
}{\sqrt{c_{1}}}\right.  \notag \\
& +\left. \frac{\tilde{\imath}_{2}-2c_{1}\tilde{\imath}_{3}+c_{1}^{2}\tilde{%
\imath}_{4}+2\gamma c_{1}\left( 1-c_{1}\right) \tilde{\imath}_{2}\left( 
\tilde{j}_{3}-2c_{1}\tilde{j}_{4}+c_{1}^{2}\tilde{j}_{5}\right) }{1-c_{1}}%
\right]  \label{e}
\end{align}%
where the integrals $\tilde{\imath}_{n}\left( \xi ,c_{1}\right) ,$ $\tilde{j}%
_{n}\left( \xi ,c_{1}\right) $ are defined in Ref. \cite{for10} and
evaluated at $\bar{\xi}\left( \mu ,c_{1}\right) $.\ This energy density is
plotted in Fig. \ref{ed}.

In the ordered phase, i.e. for $\mu \ggg \Lambda _{\text{YM}}$ where $%
g^{2}\left( \mu \right) \ll 1$, the energy density can be calculated
perturbatively (in $g^{2}$). Since fluctuations $\varphi _{<}^{a}$ around $%
U_{<}\sim 1$ are small in this phase, one may approximate $U_{<}=\exp \left(
ig\varphi _{<}^{a}\lambda ^{a}\right) =1+ig\varphi _{<}^{a}\lambda
^{a}+O\left( g^{2}\right) .$ After adding the hard-mode contribution (\ref%
{epsh}) and discarding the zero-point contribution, this results in 
\begin{equation}
\varepsilon \left( \mu ,c_{1}\right) =\frac{N_{c}^{2}-1}{4\pi ^{2}}\mu ^{4}%
\frac{1-c_{1}}{c_{1}^{2}}\left[ -\frac{c_{1}^{3}+15c_{1}^{2}-50c_{1}+30}{%
30\left( 1-c_{1}\right) ^{2}}+\frac{\mathrm{arctanh}\sqrt{c_{1}}}{\sqrt{c_{1}%
}}\right] .
\end{equation}%
It is reasonable to expect that this perturbative result remains
qualitatively reliable down to the phase transition at $\mu _{c}$ 
\cite{kog95}. The singularity of the energy density at $c_{1}\rightarrow 1$
encodes the vacuum instability for $c_{1}\geq 1$ and thus automatically
ensures that the wave functional remains normalizable during the variational
analysis.

The most important lesson of the above analysis is that $\varepsilon \left(
\mu ,c_{1}\right) $ \emph{increases} monotonically with $\mu $ and $c_{1}$
(for $-2<c_{1}<1$) in the ordered phase while the energy density (\ref{e})
in the strongly-coupled disordered phase monotonically \emph{decreases} with 
$\mu $ and $c_{1}$, up to the phase transition. This indicates that the
vacuum energy density becomes minimal at the phase boundary in the
disordered phase, i.e. at $\bar{\xi}=0_{+}$ (where the number of massless
particles becomes maximal \cite{for10}). The precise minimum, $\bar{%
\varepsilon}\left( \mu ^{\ast },c_{1}^{\ast }\right) \simeq -210.59\Lambda _{%
\text{YM}}^{4},$ is reached at $c_{1}^{\ast }\simeq 0.15$ with $\mu ^{\ast
}=\mu _{c}\left( c_{1}^{\ast }\right) =8.61\Lambda _{\text{YM}}.$
These values justify the perturbative treatment of\ the hard modes and of
the $4U$ contributions. The $c_{1}$ corrections reduce the vacuum energy
density by about 11\% and provide a rather substantial improvement of the
wave functional.

\subsection{Gluon condensate and quasigluon kinetic mass}

At the physical parameter values, i.e. at the border of the disordered phase
where the energy is minimal, the gluon condensate becomes 
\begin{equation}
\left\langle F^{2}\right\rangle =-\frac{N_{c}^{2}-1}{\pi ^{2}}\mu ^{4}\left[ 
\frac{7c_{1}^{3}-20\gamma ^{\ast }c_{1}^{3}+15c_{1}^{2}+20\gamma ^{\ast
}c_{1}^{2}-50c_{1}+30}{30c_{1}^{2}\left( 1-c_{1}\right) }-\left( \frac{%
1-c_{1}}{c_{1}^{2}}+\frac{2\gamma ^{\ast }}{3}\right) \frac{\mathrm{arctanh}%
\sqrt{c_{1}}}{\sqrt{c_{1}}}\right]  \label{f}
\end{equation}%
($\gamma ^{\ast }=g^{2}\left( \mu ^{\ast }\right) N_{c}/\pi ^{2}\simeq 1.012$%
). Numerically, this implies 
\begin{equation}
\left\langle \frac{\alpha }{\pi }F^{2}\right\rangle =20.87\Lambda _{\text{YM}%
}^{4}\simeq 0.011\text{ GeV}^{4}  \label{gc}
\end{equation}%
(for $\Lambda _{\text{YM}}\simeq 0.15$ GeV), i.e. an about 25\% larger value
than in the uncorrected $c_{1}=0$ case. The result (\ref{gc}) lies
comfortably within the standard range $\left\langle \left( \alpha /\pi
\right) F^{2}\right\rangle =0.0080-0.024$ GeV$^{4}$ obtained from QCD sum
rules \cite{for05}.

Our finite and positive result for $c_{1}$ has further interesting
consequences since it reshapes the composition and dispersion of the vacuum
field population. Indeed, the attractive IR interactions generated by $%
c_{1}>0$ deplete the density of ultralong-wavelength $k\rightarrow 0$ modes
and populate the $k\sim \mu $ modes more strongly. This is consistent with
the expected average wavelength $\lambda \sim \Lambda _{\text{YM}}^{-1}$ of
the vacuum fields. Since $G^{-1}\left( k\right) $ describes the dispersion
relation $\omega \left( k\right) $ of \textquotedblleft
quasigluon\textquotedblright\ modes\ in the vacuum, furthermore, one may
relate $c_{1}$ to the modulus of the dimensionless quasigluon group velocity 
$\vec{v}\left( \vec{k}\right) =\partial G_{<}^{-1}\left( \vec{k}\right)
/\partial \vec{k}$ at $k=\mu $ \cite{for10},%
\begin{equation}
\left\vert c_{1}\right\vert =\frac{v\left( \mu \right) }{v\left( \mu \right)
+2}.
\end{equation}%
For $0>c_{1}>1$ (as in our case), furthermore, the \textquotedblleft
effective \emph{kinetic} gluon mass\textquotedblright\ $\overline{m}_{\text{g%
}}$, which relates velocity and momentum as $\vec{k}=\overline{m}_{\text{g}}%
\vec{v}$, is negative. Hence $\vec{v}$ is opposite to the momentum, causing
the \textquotedblleft quasigluons\textquotedblright\ in the vacuum to
decelerate when an external force is applied. (Such dispersions are
encountered in several condensed-matter systems and are in stark contrast to
the behavior of free gluons.) Hence quasigluons (with their small scattering
amplitudes) may show a negative differential color resistance.

\section{Infrared degrees of freedom}

\label{IRSPs}

Our above representation of the vacuum dynamics in terms of the \emph{%
gauge-invariant} low-energy fields $U_{<}$ provides the opportunity to
search for specific $U_{<}$ which may play a particularly important or even
dominant role in the generating functional (and hence universally in all
low-energy amplitudes) \cite{cor11}. If such fields exist, they can be
regarded as universal infrared degrees of freedom (IRdofs). In contrast to
other proposed IRdof candidates (e.g. classical gauge-field solutions like
instantons \cite{sch98}, or monopole and vortex configurations), the IRdofs
expressed in terms of $U_{<}$ are gauge invariant and contain crucial
quantum effects (e.g. those which stabilize the instanton size, see below).
From a practical perspective, these IRdofs will be useful as well since many
technical problems encountered when dealing with gauge-dependent fields are
avoided from the outset. Below we will show that large classes of such
IRdofs indeed exist and review how their stability and topology emerges. We then
construct important IRdof classes explicitly and discuss their properties and
physical interpretation.

\subsection{Gauge-invariant saddle point expansion}

\label{spex}

We start from the vacuum overlap matrix element, i.e. the functional
integral 
\begin{equation}
Z=\int DU_{<}\exp \left( -\Gamma \left[ U_{<}\right] \right)  \label{zsoft}
\end{equation}%
over the soft modes, with the action $\Gamma $ given by Eq. (\ref{gsm0}).\
(Sources can be included when needed.)\ A steepest descent\ approximation
for $Z$ can be set up by expanding $U_{<}$ around the\ saddle point fields $%
\bar{U}_{i}\left( \vec{x}\right) $, i.e. the local minima of the soft-mode
action (\ref{gsm0}) which solve%
\begin{equation}
\left. \frac{\delta \Gamma \left[ U_{<}\right] }{\delta U_{<}\left( \vec{x}%
\right) }\right\vert _{U_{<}=\bar{U}_{i}^{\left( Q\right) }}=0.
\label{spaeq}
\end{equation}%
(Different topological charges (see below) are summarily denoted by $Q$
since the action is varied in each topological sector separately.) To
leading order, the saddle point expansion for $Z$ is then a weighted sum (or
integral -- the symbolic label $i$ becomes continuous when the saddle points
form continuous families) over the contributions from all relevant solutions 
$\bar{U}_{i}^{\left( Q\right) }$, 
\begin{equation}
Z\simeq \sum_{Q\in Z,i}F_{i}\left[ \bar{U}_{i}^{\left( Q\right) }\right]
\exp \left( -\Gamma \left[ \bar{U}_{i}^{\left( Q\right) }\right] \right) ,
\label{zspa}
\end{equation}%
where nontrivial pre-exponential factors $F_{i}$ are typically generated by
zero-mode contributions.

For the general analysis and explicit solution of Eq. (\ref{spaeq}) we adopt
the parametrization $U_{<}\left( \vec{x}\right) =\exp \left[ \phi \left( 
\vec{x}\right) \hat{n}^{a}\left( \vec{x}\right) \lambda ^{a}/\left(
2i\right) \right] $ of the SU$\left( N_{c}\right) $ elements and work
directly with the $N_{c}^{2}-1$ independent degrees of freedom of $U_{<}$,
i.e. the unit vector field $\hat{n}^{a}$ and the spin-0 field $\phi $. For
simplicity, we will also specialize to $N_{c}=2$ and use the first two terms
in the expansion (\ref{gm1}) of the inverse finite-mass gluon propagator $%
G^{-1}\left( k\right) =\sqrt{k^{2}+\mu ^{2}}$ as a template for the
covariance \cite{for06}. The soft-mode Lagrangian can can then be written as
a sum of two- and four-derivative terms, 
\begin{equation}
\mathcal{L}\left( U_{<}\right) =\mathcal{L}_{2d}\left( \phi ,\hat{n}\right) +%
\mathcal{L}_{4d}\left( \phi ,\hat{n}\right) .  \label{l24d}
\end{equation}%
(For the explicit expressions see Ref. \cite{for06}.) The saddle point
equation (\ref{spaeq}), when specialized to variations\ with respect to $%
\phi $ and $\hat{n}^{a}$, becomes a system of four nonlinear partial
differential equations. Its localized solutions can be shown to be stable
under scale transformations, due to the virial theorem $\Gamma _{2d}\left(
1\right) =\Gamma _{4d}\left( 1\right) $ \cite{for06}. (Clearly the
four-derivative term $\Gamma _{4d}$ is crucial here -- truncation of the
gradient expansion (\ref{gm1}) to two powers of $\partial U_{<}/\mu $ is
therefore the minimal approximation which supports stable saddle points.)\
The origin of this stability can be traced to the mass scale $\mu $ emerging
from the out-integrated short-wavelength quantum fluctuations.

An already mentioned, crucial benefit of the gauge-projected wave
functionals (\ref{ginvvwf}) is that they fully implement the nontrivial
topology of the gauge group and fields. The $U_{<}$ fields thereby inherit
three integer topological quantum numbers \cite{for06}: a winding number $Q%
\left[ U_{<}\right] $ (characterizing the homotopy group $\pi _{3}\left(
S^{3}\right) =Z$), a monopole-type degree $q_{m}\left[ \hat{n}\right] $
based on $\pi _{2}\left( S^{2}\right) =Z$ and finally a linking number $q_{H}%
\left[ \hat{n}\right] $ in the Hopf bundle $\pi _{3}\left( S^{2}\right) =Z$
which classifies knot solutions. This topology entails two lower action
bounds \cite{for06} of Bogomol'nyi type,%
\begin{equation}
\Gamma \left[ U_{<}\right] \geq \frac{12\pi ^{2}}{g^{2}\left( \mu \right) }%
\left\vert Q\left[ U_{<}\right] \right\vert ,\text{\ \ \ \ \ }\Gamma \left[
\phi _{k}=\left( 2k+1\right) \pi ,\hat{n}\right] \geq \frac{%
2^{9/2}3^{3/8}\pi ^{2}}{g^{2}\left( \mu \right) }\left\vert q_{H}\left[ \hat{%
n}\right] \right\vert ^{3/4},  \label{bb}
\end{equation}%
which ensure that contributions to soft amplitudes from saddle points in
high charge sectors can generally be neglected. This allows for practicable
truncations of the saddle-point expansion. (Saturation of the first bound
requires the fields to solve the Bogomol'nyi-type equation $\partial
_{i}L_{j}=\mp \mu \varepsilon _{ijk}L_{k}$, incidentally, which can be
considered as the analog of the self-(anti)-duality equation in Yang-Mills
theory.)

\subsection{Important examples of gauge-invariant infrared degrees of freedom%
}

In general, the saddle-point solutions have to be found numerically. Among
the exceptions are the translationally invariant vacuum solutions $%
U_{c}=const.$ (which are the absolute action minima $\Gamma \left[ U_{c}%
\right] =0$) and several nontrivial solution classes which can be found
analytically. In addition, important and sufficiently symmetric solutions
classes can often be obtained by solving substantially simplified field
equations \cite{for06}. (The typically smaller action values\ of solutions
with higher symmetry generate a stronger impact on the matrix elements,
furthermore.)

As an example for nontrivial analytical solutions, we consider $U_{<}$
fields with constant $\hat{n}^{a}$ for which the saddle point equation
becomes linear: $\partial ^{2}\left( \partial ^{2}\phi -2\mu ^{2}\phi
\right) =0.$ The general solution does not carry any topological charge and
was found in Ref. \cite{for06}. The subset of spherically symmetric
solutions with finite action, in particular, is%
\begin{equation}
\bar{\phi}^{\left( \hat{n}=c\right) }\left( r\right) =c_{1}+\frac{c_{2}}{%
\sqrt{2}\mu r}\left( 1-e^{-\sqrt{2}\mu r}\right)  \label{phineqc}
\end{equation}%
with the action $\Gamma \left[ \phi ^{\left( \hat{n}=c\right) },\hat{n}_{c}%
\right] =c_{2}^{2}\pi /\left( \sqrt{2}g^{2}\left( \mu \right) \right) $.
Since Eq. (\ref{phineqc}) is not subject to topological bounds, it
continuously turns into one of the vacuum solutions for $c_{2}\rightarrow 0$.

A particularly important saddle-point solution class consists of topological
solitons of \textquotedblleft hedgehog\textquotedblright\ type, 
\begin{equation}
\hat{n}^{a}\left( \vec{x}\right) =\hat{x}^{a},\text{ \ \ \ \ \ }\phi \left( 
\vec{x}\right) =\phi ^{\left( hh\right) }\left( r\right)  \label{hh}
\end{equation}%
($\hat{x}^{a}\equiv \vec{x}/r$, $r\equiv \left\vert \vec{x}\right\vert $).
Well-defined hedgehog fields must satisfy the boundary condition $\phi
^{\left( hh\right) }\left( 0\right) =2k_{1}\pi $ (regularity at the origin
further requires $\phi ^{\prime \prime }\left( 0\right) =0$) and
finite-action fields additionally have $\phi ^{\left( hh\right) }\left(
\infty \right) =2k_{2}\pi $ where $k_{1,2}$ and the charge $Q=k_{1}-k_{2}$
are integers. The more general boundary conditions 
\begin{equation}
\phi ^{\left( hh\right) }\left( 0\right) =n\pi ,\text{ \ \ \ \ }\phi
^{\left( hh\right) }\left( \infty \right) =m\pi ,\text{ \ \ \ \ }Q\left[
\phi ^{\left( hh\right) }\right] =\frac{n-m}{2}  \label{q-hh}
\end{equation}%
($n,m$ integer) additionally admit infinite-action solutions with half-integer
winding numbers $Q$ (for either $m$ or $n$ odd). All hedgehog fields further
carry the monopole-type charge $q_{m}^{\left( hh\right) }:=q_{m}\left[ \hat{x%
}\right] =\pm 1$. Due to the periodicity in $\phi $, it is sufficient to
consider boundary values in the range $\phi \left( 0\right) \in \left]
0,2\pi \right] $. The dynamics of $\phi \left( r\right) $ is governed by the
radial Lagrangian 
\begin{equation}
\mathcal{L}^{\left( hh\right) }\left( r\right) =\frac{\pi }{g^{2}\left( \mu
\right) \mu }\left[ \frac{1}{2}\left( r\phi ^{\prime \prime }\right)
^{2}+\left( 3+\mu ^{2}r^{2}\right) \left( \phi ^{\prime }\right) ^{2}+4\mu
^{2}\left( 1-\cos \phi \right) \right] .  \label{lrad}
\end{equation}%
The hedgehog saddle points, found numerically in Ref. \cite{for06}, turn out
to comprise mainly contributions from and around the gauge orbits of the
classical Yang-Mills solutions, i.e. (multi-) instantons and (multi-)
merons. The potential term in Eq. (\ref{lrad}) is analogous to that of a
one-dimensional pendulum in a gravitational field, with stable (unstable)
equilibrium positions at $\phi =\pi $ ($\phi =0$), modulo multiples of $2\pi 
$.

\begin{figure}[tbp]
\begin{center}
\includegraphics[width=.35\textwidth,angle=270]{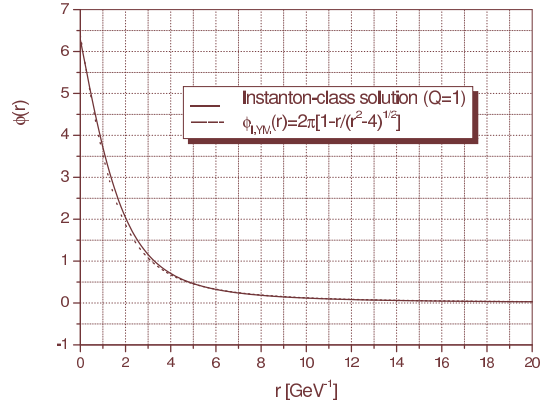}
\end{center}
\caption{The 1-instanton class solution. The dashed line corresponds to the
Yang-Mills instanton.}
\label{iclass}
\end{figure}

We first discuss the regular hedgehog solutions. Their three boundary
conditions $\phi \left( 0\right) =2\pi ,$ $\phi ^{\prime \prime }\left(
0\right) =0$ and $\phi \left( \infty \right) =2\pi \left( 1-Q\right) $ imply
that for a given $Q$ all of them can be found by varying the initial slope $%
\beta :=\phi ^{\prime }\left( 0\right) $. (For the irregular solutions with $%
\phi \left( 0\right) =\pi $ see Ref. \cite{for06}.) The regular solutions
turn out to contain one \emph{finite-action} solution for each $Q$, denoted
as the \textquotedblleft $\left\vert Q\right\vert $ (anti) instanton
class\textquotedblright , and the remaining, continuous (in $\beta $)
infinite-action families, the \textquotedblleft $2\left\vert Q\right\vert $
(anti) meron classes\textquotedblright . The 1-instanton class solution is
depicted in Fig. \ref{iclass}. Its relative gauge orientation $%
U=U_{-}^{-1}U_{+}$ is even quantitatively close to that of the Yang-Mills 
\emph{instanton} \cite{sch98} (orbit) (see dashed curve in Fig. \ref{iclass}%
). This confirms that the $\left\vert Q\right\vert $ instanton classes
indeed primarily summarize Yang-Mills instanton contributions. However, they
also contain crucial, dilatation-breaking quantum corrections which
dynamically stabilize the instanton size at about $\rho \simeq 2\mu ^{-1}$,
compatible with instanton liquid model \cite{sch98} and lattice \cite{mic95}
results, and thus overcome the chronic IR instabilities of classical
Yang-Mills instanton gases. Since instanton effects play important roles in
Yang-Mills theory (e.g. in the $\theta $ vacuum \cite{cal78} and in spin-0
glueball physics \cite{for05,sch98,for08}), it is crucial that they are (at
least partly) included in the vacuum functional (\ref{ginvvwf}). In fact,
approximate non-hedgehog solutions corresponding dominantly to ensembles of
instantons and anti-instantons should also exist and play prominent roles
(since they would be enhanced by a large \textquotedblleft
entropy\textquotedblright ,\ as in phenomenologically successful
\textquotedblleft instanton liquid\textquotedblright\ models \cite{sch98}).

All remaining regular (i.e. $\phi \left( 0\right) =2\pi $) hedgehog
solutions, with initial slopes $\beta $ between the discrete instanton-class
values $\beta _{I,Q}$, form the $2\left\vert Q\right\vert $ (anti-) meron 
classes.
Those approach one of the values $\phi _{M}\left( \infty \right) =\left(
2k+1\right) \pi $ at spacial infinity and therefore have infinite action, as
the Yang-Mills merons \cite{cal78}. Moreover, solutions with $\phi \left(
\infty \right) =\left( 2k+1\right) \pi $, corresponding to an odd number of
merons, carry the half-integer topological charge $Q$ of their Yang-Mills
meron counterparts. In addition, quantum effects ensure that our meron-class
solutions acquire a finite size and therefore remain nonsingular. Since the $%
2\left\vert Q\right\vert $-meron classes appear in continuous families
(parametrized by their \textquotedblleft size\textquotedblright\ $\beta
^{-1} $), furthermore, their large entropy will help to overcome their
infinite-action suppression in functional integrals. As in Yang-Mills
theory, merons could then play a physical role, e.g. in the confinement
mechanism \cite{cal78}.

We conclude our discussion of selected saddle-point solutions with one of
the most intriguing classes, consisting of (solitonic) links and knots.
Those emerge from a generalization of Faddeev-Niemi theory \cite{fad70},%
\begin{equation}
\mathcal{L}^{\left( \phi _{k}\right) }\left( \vec{x}\right) =\frac{\mu }{%
g^{2}\left( \mu \right) }\left[ \left( \partial _{i}\hat{n}^{a}\right) ^{2}+%
\frac{1}{\mu ^{2}}\left( \varepsilon ^{abc}\partial _{i}\hat{n}^{b}\partial
_{j}\hat{n}^{c}\right) ^{2}+\frac{1}{2\mu ^{2}}\left( \varepsilon ^{abc}\hat{%
n}^{b}\partial ^{2}\hat{n}^{c}\right) ^{2}\right] ,  \label{ln}
\end{equation}%
which is included in our soft-mode Lagrangian for constant $\phi _{k}=\left(
2k+1\right) \pi $. The corresponding solution classes $U_{<}\left( \hat{n}%
\right) $ describe twists, linked loops and knots made of closed color
fluxtubes. Since Eq. (\ref{ln}) follows uniquely from the VWF (\ref{ginvvwf}%
) and the Yang-Mills dynamics, our approach provides a new framework and
physical interpretation for such solutions. In fact, they remerge as
gauge-invariant IR degrees of freedom representing sets of gauge-field
orbits with a collective Hopf charge. While the $\hat{n}$ field of the
Faddeev-Niemi model is interpreted as a gauge-dependent local color
direction in the vacuum, in particular, our $\hat{n}$ is manifestly
gauge-invariant. This may put the tentative interpretation of such knot
solutions as glueballs on a firmer basis.

\section{Summary and conclusions}

We have studied gauge-invariant wave functionals for the Yang-Mills vacuum
which incorporate asymptotic freedom and an \emph{a priori} general
dispersion for the infrared gluons in Gaussian core functionals. In this at
present probably richest analytically manageable and gauge-invariant trial
functional basis, we have then variationally determined several vacuum
properties. Dimensional transmutation, dynamical mass generation and gluon
condensation emerge transparently and generate mass scales consistent with
other approaches. In addition, the improved vacuum description in the
infrared predicts a negative \emph{kinetic} mass of the soft gauge-field
modes and thus suggests a negative differential color resistance of the
Yang-Mills vacuum.

Another benefit of the gauge-invariant framework is that the dynamics can be
reformulated as an effective theory which represents sets of gluon orbits as
gauge-invariant matrix fields subject to higher-gradient interactions. In
this effective theory we have set up a saddle-point expansion to determine
the collective fields with maximal impact on functional integrals. These
saddle points play the role of gauge-invariant infrared degrees of freedom. They
are stabilized by the dynamical mass generation mechanism and inherit a rich
topological structure (three topological charges and action bounds of
Bogomol'nyi type) from the Yang-Mills gauge group. Moreover, they provide
the principal input for a systematic saddle-point expansion of soft
amplitudes (such as glueball correlators).

Several of the more symmetric and important saddle-point solution classes
have been found explicitly. Among them are topological solitons related to
the classical Yang-Mills (multi-) pseudoparticle solutions which mediate
tunneling processes in the vacuum. Those generate a gauge-invariant
representation of instanton and meron effects which includes
quantum fluctuations. The latter stabilize the pseudoparticle sizes
dynamically, in particular, and thereby cure the notorious infrared deseases
encountered in dilute instanton ensembles. Similar solutions with other
types of topological charges exist as well but seem to have no obvious
counterparts in classical Yang-Mills theory. One of the most intriguing
solution classes, finally, consists of solitonic links and knots. Those
emerge from a generalization of Faddeev-Niemi theory which turns out to be
embedded in our soft-mode dynamics. Hence in our framework the knot
solutions find a new and in particular gauge-invariant physical
interpretation, potentially related to glueballs.

\acknowledgments{It is a pleasure to thank the organisers for a very 
relaxed and stimulating workshop.}

\end{document}